\documentclass[conference]{IEEEtran}
\IEEEoverridecommandlockouts

\usepackage{cite}
\usepackage{amsmath,amssymb,amsfonts}
\usepackage{algorithmic}
\usepackage{graphicx}
\usepackage{textcomp}
\usepackage{xcolor}
\usepackage{hyperref}  
\usepackage{array} 
\usepackage{titlesec}
\usepackage{afterpage}
\usepackage{lineno}
\usepackage{array} 
\usepackage{adjustbox}
\usepackage{multirow} 
\usepackage{bigstrut} 

\def\BibTeX{{\rm B\kern-.05em{\sc i\kern-.025em b}\kern-.08em
    T\kern-.1667em\lower.7ex\hbox{E}\kern-.125emX}}
\begin{document}

\title{Pilot Study on Generative AI and Critical Thinking in Higher Education Classrooms}

\author{\IEEEauthorblockN{William Franz Lamberti\IEEEauthorrefmark{1},
Samantha Rose Lawrence\IEEEauthorrefmark{2}, Dominic White\IEEEauthorrefmark{3},
Sunbin Kim\IEEEauthorrefmark{4}, and Sharmin Abdullah\IEEEauthorrefmark{5}}
\IEEEauthorblockA{\textit{Computational and Data Sciences}\\
\textit{College of Science}\\
\textit{George Mason University }\\
Fairfax, VA, United States \\
Email: \IEEEauthorrefmark{1}wlamber2@gmu.edu,
\IEEEauthorrefmark{2}slawre2@gmu.edu,
\IEEEauthorrefmark{3}dwhite34@gmu.edu,
\IEEEauthorrefmark{4}skim253@gmu.edu,
\IEEEauthorrefmark{5}sabdul9@gmu.edu}}

\maketitle
\IEEEpubidadjcol

\begin{abstract}
 Generative AI (GAI) tools have seen rapid adoption in educational settings, yet their role in fostering critical thinking remains underexplored. While previous studies have examined GAI as a tutor for specific lessons or as a tool for completing assignments, few have addressed how students critically evaluate the accuracy and appropriateness of GAI-generated responses. This pilot study investigates students' ability to apply structured critical thinking when assessing Generative AI outputs in introductory Computational and Data Science courses. Given that GAI tools often produce contextually flawed or factually incorrect answers, we designed learning activities that require students to analyze, critique, and revise AI-generated solutions. Our findings offer initial insights into students' ability to engage critically with GAI content and lay the groundwork for more comprehensive studies in future semesters. 
\end{abstract}


\section{Introduction}

Artificial Intelligence (AI) has been in use in education since the 1960s \cite{doroudi2023intertwined}.  Over the past several years, usage of Generative Artificial Intelligence (GAI) has gained vast prevalence within society, including higher-education settings \cite{zawacki2019systematic, chiu2023systematic}.  The number of AI-associated educational programs has tripled since 2017\cite{maslej_artificial_2024}, and instructors have struggled to keep up with the rapid pace of AI advancements\cite{kitcharoen2024enhancing}. According to the Artificial Intelligence Index report, 63\% of K-12 teachers use ChatGPT for their work\cite{maslej_artificial_2024}. With recent investments by large private companies such as Microsoft, OpenAI, and Anthropic into a nationwide initiative that trains teachers on how to incorporate AI into our curriculum, it is plausible that GAI will continue to occupy a place in higher education\cite{nash1, microsoftannouncement}. 

Throughout human history, technological advancements in education have brought skepticism and concern. Notably in Plato's \textit{Phaedrus}, Socrates expresses cynicism at the invention of writing systems\cite{plato2022lysis}. Since Descartes first conceptualization of intelligent machines, which he entitled automata in 1637\cite{discoursdescartes, cdi_proquest_ebookcentralchapters_3338771_16_318}, questions have abounded at the possibility of nonhuman intelligence. This period also marked the beginning of an ambivalence towards technologies that replicate aspects of human behavior\cite{amsterdam_automata}. That ambivalence remains today, with the most recent Pew Research Center polls stating 35\% of U.S. adults believe that AI will have a negative impact on America in the next 20 years, and 43\% of U.S. adults believe increased usage of AI is more likely to harm them\cite{pew_opinion_citation}. Just as persistent public skepticism reflects concerns of broad impacts on society, researchers have already begun to investigate the impact GAI has on education. 

Research investigating the impact of classroom useage of GAI on students critical thinking is limited, both in volume and study design\cite{premkumar2024impact}. Students who use GAI are more likely to procrastinate and less likely to remember what they learned\cite{abbas2024harmful}.  However, when used to complement curricula GAI has been shown to increase critical thinking skills, while reducing the need for prior knowledge on the subject\cite{zhao2025generative, cong2025critical}. 

Although GAI use improves classroom efficiency in tasks that are easily automated, such as reviewing work and providing feedback\cite{srinivasa2022harnessing}, with 88\% of K-12 teachers believe ChatGPT will have a positive impact on education\cite{maslej_artificial_2024}, educators are cautious before bringing AI into the classroom. Research on ethical applications of AI are minimal and a limited number of studies provide in-depth analysis of risks and concerns AI usage poses in education\cite{zawacki2019systematic}. Student AI literacy is influenced by both controllable and uncontrollable factors, such as country of origin, gender,  discipline of study, and previous courses taken\cite{o2024factors}.  

GAI has shown to be beneficial in many educational contexts, such as skill development, personalized learning, research aggregation, and supporting equitable learning experiences\cite{kasneci2023chatgpt}. Usage of GAI technology shows benefit to educators by improving teacher engagement in student feedback by 13\% \cite{demszky2024can}, assisting in curriculum adaptation\cite{karatacs2025reshaping}, and high potential for automating aspects of the grading process\cite{floden2025grading, jukiewiczfuture}. 

It is well known that GAI do not consistently give accurate responses\cite{dahlkemper2023physics}, nor have the ability to provide precise output when given a nuanced prompt requiring context \cite{gravel2023learning}. Additionally, using GAI to aid academic work, such as writing essays, has been associated with reduced levels of brain activity-hinting at potential negative neurological impacts of long-term use \cite{kosmyna2025your}.  The integration of GAI into daily life has raised concerns and skepticism for many\cite{oprea2024skepticism}, and such doubt may be beneficial, as it is suggested that higher acceptance of GAI output is associated with less critical thinking \cite{lee2025impact}. 

Large Language Models (LLM)\cite{alafnan2023chatgpt}, which are a branch of GAI specializing in text generation, have demonstrated capacity for coherent writing at a high level. Even old models, such as GPT-3, partially authored a research paper discussing the benefits and drawbacks of LLM usage in education\cite{cotton2024chatting} and another article was entirely written by the model, discussion and analysis of itself\cite{generativepretrainedtransformer:hal-03701250}. Connecting this to the students using GAI in classroom, rather than doing assignments with critical thinking, there is a high possibility that students might just ask assignment questions to GAI, copy and paste the AI-generated answers. In addition, students are less likely to conduct their own research when shown AI summaries\cite{pew_google_summary_citation}. Marked differences have been observed in the brains of those who use GAI to write essays, and individuals who use GAI fail to recall what they wrote\cite{kosmyna2025your}. Accordingly, we are concerned that if students rely too heavily on GAI, it may lead to the weakening of critical thinking and a reduced ability in problem-solving skills\cite{lee2025impact}.

Despite the rapid rise of GAI usage in education, there is a paucity of research on how these tools influence and impact students' learning experiences and performance. It is essential to empirically examine how these tools impact students in real-world academic settings, as usage of GAI is continually increasing\cite{pew_chatgpt_useage_citation}. The amount of US adults who have used ChatGPT has doubled since 2023, with the largest share of users being young adults. Presently, global usage of ChatGPT on a weekly basis is about 36\% \cite{loewen_global_nodate}.  Furthermore, about 33\% and 28\% are using ChatGPT for their jobs and education, respectively\cite{loewen_global_nodate}.  Furthermore, previous work has investigated teaching effectiveness of GAI in college and high school courses\cite{chan2023students}. Comparatively little has been done on how students evaluate GAI outputs. 

In response to the evolving landscape, this pilot study investigated the impact of GAI on students’ critical thinking skills, paving the way for a comprehensive future study. We have created targeted lesson plans incorporating GAI and have measured their effect on student learning outcomes within a preliminary scope. By understanding how these tools influence thinking and comprehension, educators can better integrate GAI into teaching strategies that foster deeper learning and academic integrity.

\section{Methods}

We provide some context for AI, GAI, and the statistical methods utilized in this paper.  We also provide some details of the experimental design and selection of participants for our pilot study. 

\subsection{Context for AI}

It  is important to establish clear terminology to ensure a shared understanding of the concepts discussed in this study. In particular, we focus on two key terms: AI and GAI. Although there is no unified consensus on these definition\cite{thierer_artificial_2017}, and others may use different interpretations, we adopt the definitions and descriptions outlined previously.  Specifically,  ``\textbf{AI} is defined as an operation that performed by a computer that could be performed by a human.... (\textbf{GAI}) focuses on the creation of content with a spatial component."\cite{lamberti_artificial_2024}.  

\subsection{Statistical Methods}

Statistical tests are still used in a variety of research scenarios to better understand the world around us.  It is based upon the Frequentist perspective, where the parameter in question is fixed.  We typically have a statistic, such as the sample mean, to estimate the population parameter, such as the population mean \cite{bhattacharyya_statistical_1977}. For example, the formal setup for a two-tailed hypothesis test for evaluating a population mean is \begin{equation}\label{eq: null}
    H_0: \theta=b
\end{equation}

\begin{equation}\label{eq: alt}
    H_1: \theta\neq b
\end{equation}

\noindent where $H_0$ is the null hypothesis, $H_1$ is the alternative hypothesis, $\theta$ is the population parameter, and $b$ is the value to test if $\theta$ is \cite{bhattacharyya_statistical_1977}. Additional details on the one-sided hypothesis tests can be found elsewhere \cite{bhattacharyya_statistical_1977}.

Furthermore, many of the methods presented rely upon the normal distribution.  The normal distribution is observed across many fields.  The Normal distribution relies upon two parameters: the mean (usually denoted as $\mu$) and the variance or standard deviation (usually denoted as $\sigma^2$ or $\sigma$, respectively)\cite{bhattacharyya_statistical_1977}.  The equation for the normal distribution is \begin{equation}
    \varphi_{\mu, \sigma^2}(X) = \frac{1}{\sqrt{2\pi\sigma^2}}e^{-\frac{(X-\mu)^2}{2\sigma^2}}
\end{equation}  

When collecting data, researchers typically migrate from the normal distribution to Student's $t$-distribution.  The $t$-distribution is preferred in smaller data applications since it is more informative due to the relatively heavier tails\cite{bhattacharyya_statistical_1977}.   

One of the fundamental hypothesis tests is the $z$-test which is related to the population mean \cite{bhattacharyya_statistical_1977}. Using the formulation in Equations \ref{eq: null} and \ref{eq: alt}, the null and alternative hypotheses for a two tailed test are, respectively:

\begin{equation}\label{eq: norm_null}
    H_0: \mu = b
\end{equation}

\begin{equation}\label{eq: norm_alt}
    H_1: \mu \neq b
\end{equation}

\noindent where $\mu$ is the population mean.  The $z$-tests assume that the data is normally distributed \cite{bhattacharyya_statistical_1977}.

Typically, researchers use the $t$-test instead of the $z$-test since the $t$-distribution is more accurate for smaller samples that follow a normal distribution \cite{bhattacharyya_statistical_1977}.  The formulation of these tests are exactly the same as exemplified in Equations \ref{eq: norm_null} and \ref{eq: norm_alt}.  

Linear models are important to help understand Analysis of Variance (ANOVA).  OLS defines a linear relationship between the explanatory, $X$, and response variables, $Y$ \cite{mendenhall_second_2011}.  Using matrix notation, this would be\begin{equation}\label{eq: mat}
    Y = \beta X.
\end{equation}

\noindent The parameters to be estimated are $\beta$. A normality assumption is usually applied for the errors or residuals of the OLS model \cite{lamberti_william_franz_overview_2022}. 

Equation \ref{eq: mat} is equivalent to the following:\begin{equation}\label{eq: reg}
    Y = \hat{\beta}_0 + \hat{\beta}_1X_1+\hat{\beta}_2X_2 + \dots + \hat{\beta}_jX_j + \epsilon
\end{equation}

\noindent where $Y$ is the response variable, $\hat{\beta}_0$ is the estimated intercept, $\hat{\beta}_j$ is the $j^{\text{th}}$ estimated coefficient value, $X_j$ is the $j^{\text{th}}$ independent variable, and $\epsilon$ is our error term \cite{draper_applied_1998, mendenhall_second_2011}.

ANOVA is a special case of OLS\cite{draper_applied_1998}.  The equation is exactly the same as \ref{eq: reg}. However, the definitions can be more precise.  In this setup, we have that:

\[X_j=\begin{cases}
\text{Treatment $j$ is applied}, & \text{if $X_j=1$}\\
\text{Treatment $j$ is not applied}, & \text{if $X_j=0$}
\end{cases}.
\]

\noindent In other words, all of our $X_j$'s become a series of dummy or binary variables\cite{draper_applied_1998}. 

In the case where Normality is not held, we can use the nonparametic version of ANOVA called the Kruskall-Wallis (KW) test\cite{bhattacharyya_statistical_1977}.  The setup for the KW test is:

\begin{multline}\label{eq: kw_null}
    H_0: \text{All $k$ continuous population} \\ \text{distributions are the same}
\end{multline}

\begin{equation}\label{eq: kw_alt}
    H_1:  \text{At least one differs}
\end{equation}

\noindent where $k$ is the number of distributions were are comparing.  For example, if we have 1 treatment and 1 control, we would have that $k=2$.  

Typically, the goal of ANOVA is to test if different treatments impact a given response.  In medicine, one might test a new drug against a placebo, and the response would be the given medical outcome (i.e., blood pressure readings).  In education, one might test a new lesson plan against no changes, and the response would be the resulting end of class quiz grades. 

In our case, the treatment is a prerecorded video and a homework assignment.  The control is the same homework assignment.  While both CDS 101 and CDS 130 are related to critical thinking and GAI, each course had slightly different videos and assignments.  The details of the assignments and videos are provided in the Appendix. 

\subsection{Experimental Design}

We employed a quasi-experimental design with one treatment and one control section of two data science courses. The independent variable was a 15 minute lecture on GAI output. The dependent variables are provided in Table \ref{tab:vars}.

\begin{table}
    \centering
    \begin{tabular}{c|c}
        Class & GAI Assignment Name  \\\hline 
        CDS 101 & Generative AI Assignment \\
        CDS 130 & WA 13.5 \\ \hline \\
    \end{tabular}
    \caption{Summary table of GAI assignment names in different courses. }
    \label{tab:vars}
\end{table}

CDS 130 is an introductory computing class within the Computational \& Data Sciences department at George Mason University, as well as a Mason Core class. Mason Core is a foundational program which instills knowledge and skills in all Mason Graduates\cite{gmucommoncore}. CDS 130 serves a substantial percentage of the undergraduate student population, as approximately 1,000 students take this course yearly. 
CDS 130 had 4 sections: 2 taught in Summer Session A and 2 taught in Summer Session C.  

CDS 101 is a large introductory course within the Computational and Data Sciences department, with an accompanying laboratory section entitled CDS 102. Together, these courses teach students how to use computers to analyze, manipulate, and visualize data.  Approximately 700 students take this course yearly.
CDS 101 had two sections taught asynchronously in Summer Session A and Summer Session C. 

\subsection{Selection of Participants}

Participants were from 6 CDS classes offered at GMU during the Summer 2025 session.  These courses are GMU Common Core Courses \cite{gmucommoncore}, which are mandatory, foundational liberal arts courses intended to foster lifelong engagement and critical thinking skills in each student.

Student participation in the study was not required for taking the class.  Those that wished to not participate in the class were not treated differently nor removed from the class.  Additional Institutional Review Board (IRB) details are provided in the Appendix.  

The following tables provides some high level information regarding the composition of students that agreed to participate in the study.

\begin{table*}[ht!]
\centering
\begin{tabular}{|l|l|l|l|l|l|l|}
\hline
\textbf{Course \#} & \textbf{Section} & \textbf{Instructor} & \textbf{Session} & T/Con & \textbf{\# of Enrolled} & \textbf{\# of Participating}  \\
\hline
CDS 130 & A02 & Lamberti & A & T       & 5  & 1\\
CDS 130 & A03 & Lamberti & A & Con     & 16 & 4 \\ \\
CDS 130 & C01 & Abdullah & C & T       & 8  & 1\\
CDS 130 & C03 & Abdullah & C & Con     & 12 & 5 \\ \\\hline 
CDS 101 & A01 & White    & A & T       & 17 & 6\\
CDS 101 & C01 & White    & C & Con     & 16 & 4\\\hline
        &     &          &   & Totals  & 74 & 21\\
\hline
\end{tabular}
\caption{Enrollment and treatment (T) vs. control (Con) assignment. }
\label{tab:course_enrollment}
\end{table*}

\section{Results}

\subsection{Participation Rates}

As shown in Table \ref{tab:course_enrollment}, the CDS 130 ranged from 12.5\% to about 41.6\%.  If we combine all of the CDS 130 sections, we obtain a participation rate of about 27\%.  The CDS 101 A01 and C01 participation rates were about 35\% and 25 \%, respectively.  This corresponds to an overall participation rate in all CDS courses of about 28 \%.   

\subsection{CDS 101 ANOVA}

Due to the low participation rates in the the CDS 130 treament classes, ANOVA was not performed on those sections.  Despite CDS 101 being offered in separate sessions during the Summer timeframe, we performed an ANOVA analysis on those that participated as a preliminary and naïve investigation.

\begin{figure}[h!]
    \centering
    \includegraphics[width=\linewidth]{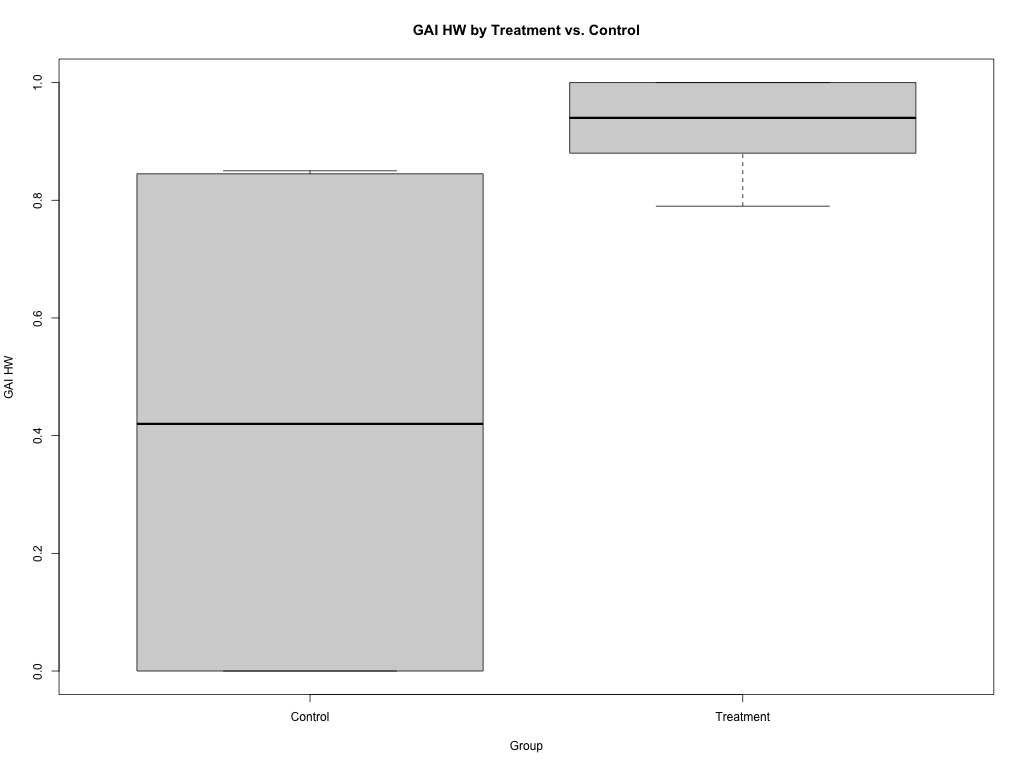}
    \caption{Whisker plot of CDS 101 Summer classes.}
    \label{fig: boxplot_cds101}
\end{figure}

The boxplots of the those in the treatment vs. control is provided in Figure \ref{fig: boxplot_cds101}.  The treatment has a large variance relative to the control.

\begin{figure}[h!]
    \centering
    \includegraphics[width=\linewidth]{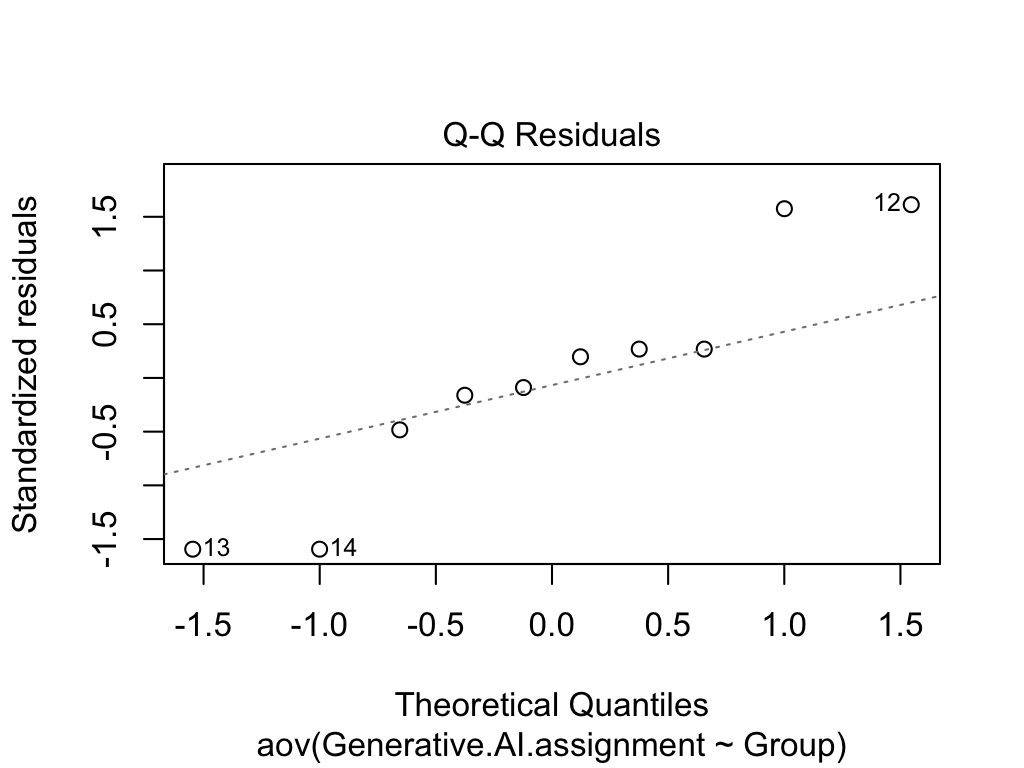}
    \caption{QQ-Plot of CDS 101 Summer classes.}
    \label{fig: qqplot_cds101}
\end{figure}

The QQ-Plot of those students in CDS 101 is provided in Figure \ref{fig: qqplot_cds101}.  There are some concerns at the tails of the QQ-Plot of some deviations from a Normal distribution.  This concern is emphasized since we are in a particularly low sample scenario.  

Thus, we performed a KW test.  Using the general setup from Equations \ref{eq: kw_null} and \ref{eq: kw_alt}, we have that 

\begin{multline}\label{eq: kw_null_101}
    H_0: \text{The treatment and control Summer 2025 distributions }\\\text{of CDS 101 for the GAI assignment are the same}
\end{multline}

\begin{equation}\label{eq: kw_alt_101}
    H_1:  \text{At least one differs}
\end{equation}

This results in the treatment being statistically significant ($p = 0.0319 $) at $\alpha = 0.05 $.  Thus, we have evidence that the video lesson impacts performance on GAI critical thinking. 

\section{Discussion}   

\subsection{Improve Study by Having Large Number of Sections}

Student willingness to participate in this study was unknown in advance. Given that participation was voluntary, we were encouraged to observe a participation rate above 10\% across all sections. However, to increase the robustness and generalizability of future results, larger class sizes or a greater number of sections would improve the likelihood of obtaining meaningful insights. Additional replications would also strengthen the findings, although these would require additional time, resources, training for new research team members, and other logistical considerations.  

\subsection{Naïve Insights from CDS 101}

We have evidence that the video lessons do have a positive impact on student outcomes for the related assignment.  As shown in the whisker plot in Figure \ref{fig: boxplot_cds101}, there are some student who are able to achieve high levels of competency without the video lesson.   If we do remove those students who earned a 0 (those students never submitted the assignment), we no longer have statistical significance.  Further, we are assuming that assigning a treament and control in different sessions adequately satisfies the experimental design.  For all of these reasons, we believe that the result of the KW test provides weak evidence.  Regardless, this provides a ``weak prior" for future studies that will be conducted in future semesters.  

\section{Conclusion}

Our pilot study on GAI and critical thinking in CDS 101 and 130 provided valuable insights for future experiments.  We now can expect about 25\% to 30\% of students in our classes to participate.  Further, we have some evidence that the lesson plan is effective for students learning critical thinking skills related to GAI.  These findings underscore the potential of integrating GAI into the curriculum, paving the way for more comprehensive studies to refine and expand upon our approach in enhancing critical thinking skills across diverse educational settings. 

\section{Acknowledgments}

The following tools were used to write this manuscript: Overleaf, Writefull (which is built into Overleaf as of 2024), Microsoft Word, Notability on an iPad, 1st generation Apple Pencil, Macbook Pro personal and work computers, Ubuntu desktop homebuilt desktop computer, Patriot AI, Zotero, Preview on Mac, Safari, and Firefox.  Tools not mentioned were unintentionally omitted.  

We would like to thank Annie Hui for her advice regarding the formulation and setup of this study.  Emilia Ermanoski in a special way, as her help with overseeing the TAs in the CDS 130 sections was instramental. We also want to thank the Stearns Center and the Computational and Data Sciences Department at GMU for their financial support.  

\section{Author Contributions}

Dr. Wlliam Franz Lamberti is the PI and corresponding author of this study. He lead the research by submitting IRB related documents, overseeing student researchers, leading other instructors, providing guidance for other CDS 130 instructors for the setup and running of the experiment, wrote the paper, edited the paper, taught ANOVA concepts to student researchers, leading funding efforts. He also ran the first two CDS 130 courses in Summer Session A.

Dr. Dominic White contributed to the research design, development of CDS 101 instructional and assessment material, data collection, and manuscript editing. He taught both summer sections of CDS 101.

Dr. Sharmin Abdullah contributed to the research by running CDS 130 classes during Summer Session C, guiding students and teaching assistants in conducting the necessary experiments and tasks for collecting data. She completed the required IRB training for this project and supported the development of the study through reviewing and editing both the manuscript and related funding proposals.

Samantha Rose Lawrence is a student researcher who conducted the literature review, wrote and edited the manuscript. She completed the required IRB training for this project. Additionally, she was a TA for both summer sections of CDS 101 and graded the assignments for CDS 101. 

Sunbin Kim is a student researcher who contributed to the foundational coding to analyze the results and to the editing of the manuscript. She also completed IRB training for this project.

\bibliographystyle{IEEEtran}
\bibliography{main.bib}

\appendix

\section*{IRB Details}

The research plan for the institutional review board (IRB) was conducted at GMU via the Office of Research Integrity and Assurance.  The review, feedback, and approval was provided via Research Administration Management Portal (RAMP).  

\section*{Details on Treatment}

Students were given a written assignment consisting of several reflective short answer prompts.
All assignments were administered uniformly across both treatment and control groups. The assignments were graded using the same rubrics, regardless of group designation. 

The treatment condition involved the integration of a 15 minute prerecorded presentation designed to teach students how to evaluate the output of generative artificial intelligence programs.

\subsection*{CDS 101 Materials}

The treatment was a 15 minute prerecorded presentation introducing LLMs, discussing concepts such as hallucinations, and expectations for GAI usage in the course. It covered two example cases, using ChatGPT to answer sample coursework questions, and explaining problems with each answer. Finally, students were given two exercises meant to complete in conjunction with the video. 

Students across all sections were given a written assignment consisting of four short answer prompts. Assignments were graded on accuracy and clarity, with the same rubric, for all sections.

\subsection*{CDS 130 Materials}

Similar to the CDS 101 materials, the treatment was about 12 minutes long.  It was a prerecorded video introducing LLMs, discussing concepts such as hallucinations, and expectations for GAI usage in the course. It covered two example cases, interpreting GAI outputs, and explaining problems with each answer. Finally, students were given two exercises meant to complete in conjunction with the video. 

Students across all sections were given a written assignment consisting of four short answer prompts. Assignments were graded on accuracy and clarity, with the same rubric, for all sections.

\section*{GitHub}

Additional resources such as the code and data used for this pilot study are available on our GitHub.  The link is \url{https://github.com/billyl320/gai_critical_thinking_pilot_gmu}

\end{document}